# Discovered a new open cluster near the poorly studied Majaess 190

A.L. Tadross

*National Research Institute of Astronomy and Geophysics, Helwan, Cairo, Egypt*

**Abstract**

We present the results of a comprehensive astrophysical study of a newly discovered open cluster, dubbed Ash-1. Using the third data release of the Gaia space mission, Gaia DR3, Ash-1 was accidentally discovered within the constellation Sagittarius in the field of the poorly studied cluster Majaess 190. Here we present the first estimates of these two clusters' primary astrophysical properties. The membership probabilities > 0.50 were assigned to both clusters using the pyUPMASK technique. The distances were determined using the parallaxes of the clusters' members, which were consistent with the isochrone fitting of the color-magnitude diagrams. The ages and distances of Majaess 190 and Ash-1 are found to be $4 \pm 0.05$ Gyr and $630 \pm 20$ Myr; 2130 and 1360 ($\pm 80$) pc, respectively. The extinctions, heliocentric distances, mass function, luminosity function, and overall masses of the studied clusters were also computed. Based on the relaxation times, it appears that both clusters are in a state of dynamic relaxation.

## 1. Introduction

A new era of precision astrometry has begun with the ESA's Gaia mission (Gaia Collaboration, 2021; Cantat-Gaudin et al., 2020) allowing us to better investigate open clusters (OCs) in our Galaxy with unprecedented accuracy and discover many of them as well. It has long been known that OCs are useful objects for studying the structure and evolution of the Milky Way Galaxy, Moraux (2016). Most of the new objects that have been identified are young and were buried in molecular clouds, and due to the high absorption, they are not visible in the optical band (Torrealba et al. 2019; Ryu & Lee 2018, Barb´a et al. 2015; Borissova et al. 2014; Bica et al. 2003).

Gaia's mission has achieved the whole sky with high precision proper motions and parallaxes which are appropriate for accurately differentiating between real clusters and field stars. As a result, there are more recent discoveries of OCs (Songmei, et al. 2023, and references therein). Bragaglia (2018) states that we can locate about 100,000 open clusters if we extend the solar vicinity to the whole galactic disc. Numerous astrophotometric studies have been carried out, and some of them have previously been discovered. Nonetheless, we anticipate that the majority of such objects' discoveries in the future will occur during the Gaia era.

The third version of the Gaia Observations (Gaia DR3) was released on June 13, 2022. It provides the position and magnitude for approximately 1.8 billion sources: coordinates ($\alpha$, $\delta$), proper motions ($\mu\alpha$, $\mu\delta$), and parallaxes ($\pi$). Moreover, the three photometric filters' magnitudes and color indices for about 1.5 billion sources. Gaia Archive can be accessed via the website (https://www.cosmos.esa.int/gaia).

*E-mail address:* altadross@nriag.sci.eg

This work is an extension of our series whose objective was to study the astrophysical properties of recently discovered, poorly understood, and/or unstudied clusters utilizing the latest missions. More than 7000 objects are currently recognized as open star clusters in the literature (Hunt & Reffert 2023 and references therein). But just a name, rough angular size, and coordinates are provided for around half of them in their citations (Kharchenko et al. 2013). One of those objects is the poorly studied cluster Majaess 190, (Majaess 2013), located in the direction of the Galactic disk, in the constellation Sagittarius. Accidentally, we discovered another OC within that field, and close to Majaess 190, which we named Ash-1 (see Sec. 2). No comprehensive studies have been carried out for those two clusters so far. So, we conducted here the first full investigation for both clusters using Gaia DR3.

The remainder of this paper is organized as follows: In Section 2, the detection of the new cluster is described. In Section 3, the data analysis is presented. Section 4 depicts the cluster's dimensions, (centers and sizes) depending on the radial density profile of each cluster. Section 5 explains the membership candidates' assessment. Section 6 carries out the analysis of the color-magnitude diagrams. Section 7 describes the mass, luminosity functions, and the dynamic state. Finally, Section 8 highlights the main conclusions, providing a list of our estimated characteristics for the studied clusters.

## 2. The new cluster's discovery

Examining the area surrounding the open star cluster Majaess 190, we have discovered another cluster entirely by chance. It exists in a denser star field than Majaess 190, so it was difficult to recognize over the background field, as shown in Fig. 1. We named Ash-1, this abbreviation accounts for the first three letters of the author's first name "Ashraf". Ash-1 appears as a clump of 410 stars in the vector point diagram, showing the co-moving stars that have similar speeds and directions in the sky. Actually, Ash-1 is found located very close to the newly discovered system CWNU-2598 (He, et al. 2023), away from it by about 0.05 deg, as shown in Fig. 1. Comparing their essential parameters, we found that they have almost similar locations, and ages, but the other parameters are somewhat different despite they fall around the same range. From this, we can conclude that the two clusters are different with some overlapping between them. So, we believe that they may represent a binary system, that needs another detailed study, see Table 1. Regardless, we will focus on Majaess 190 and Ash-1, which is our interest in the current study.

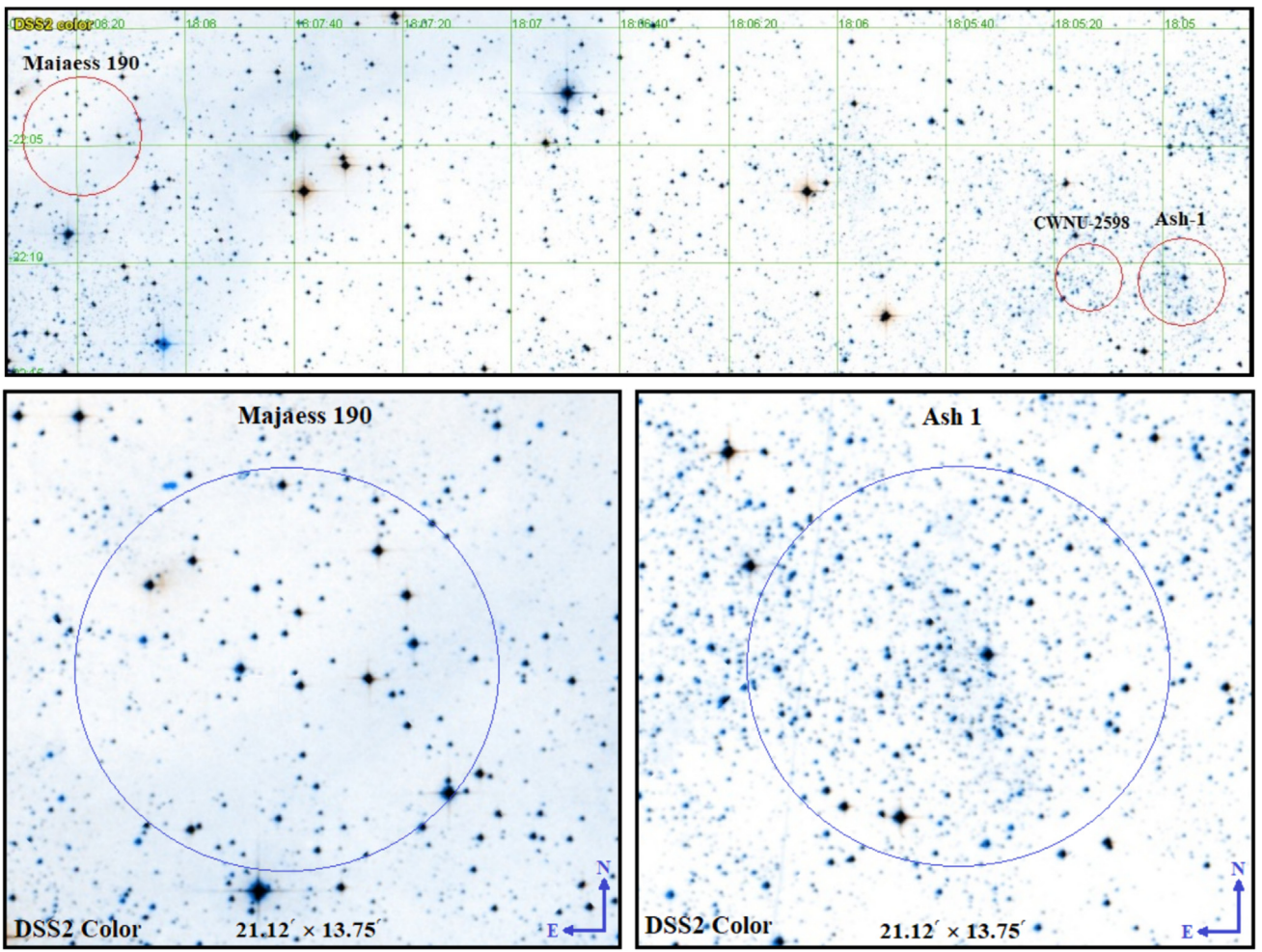

Fig. 1. The upper panel represents the locations of Majaess 190, Ash-1, and CWNU-2598 as obtained at DSS2-colored optical wavelength in ALADIN, with the equatorial coordinate grid. The lower panels focus on the two main clusters under study Majaess 190, and Ash-1. North is up, East is to the left.

Table 1
A comparison between the main astrophysical parameters of Ash-1 and CWNU-2598.

| Cluster | Glong Glat *(deg)* | Log Age *(yr)* | Plx *(mas)* | pmRA pmDE *(mas/yr)* | m-M *(mag)* | A0 *(mag)* | RV *(km/s)* | Num. stars |
|---|---|---|---|---|---|---|---|---|
| **Ash-1** | 8.03 −0.35 | 8.80 | 0.74 ± 0.07 | −0.37 ± 0.11 −2.18 ± 0.08 | 10.68 | 1.16 | −9.29 | 410 |
| **CWNU-2598** | 8.07 −0.40 | 8.55 | 0.44 ± 0.06 | −0.22 ± 0.12 −3.59 ± 0.08 | 11.40 | 5.20 | −20.58 | 67 |

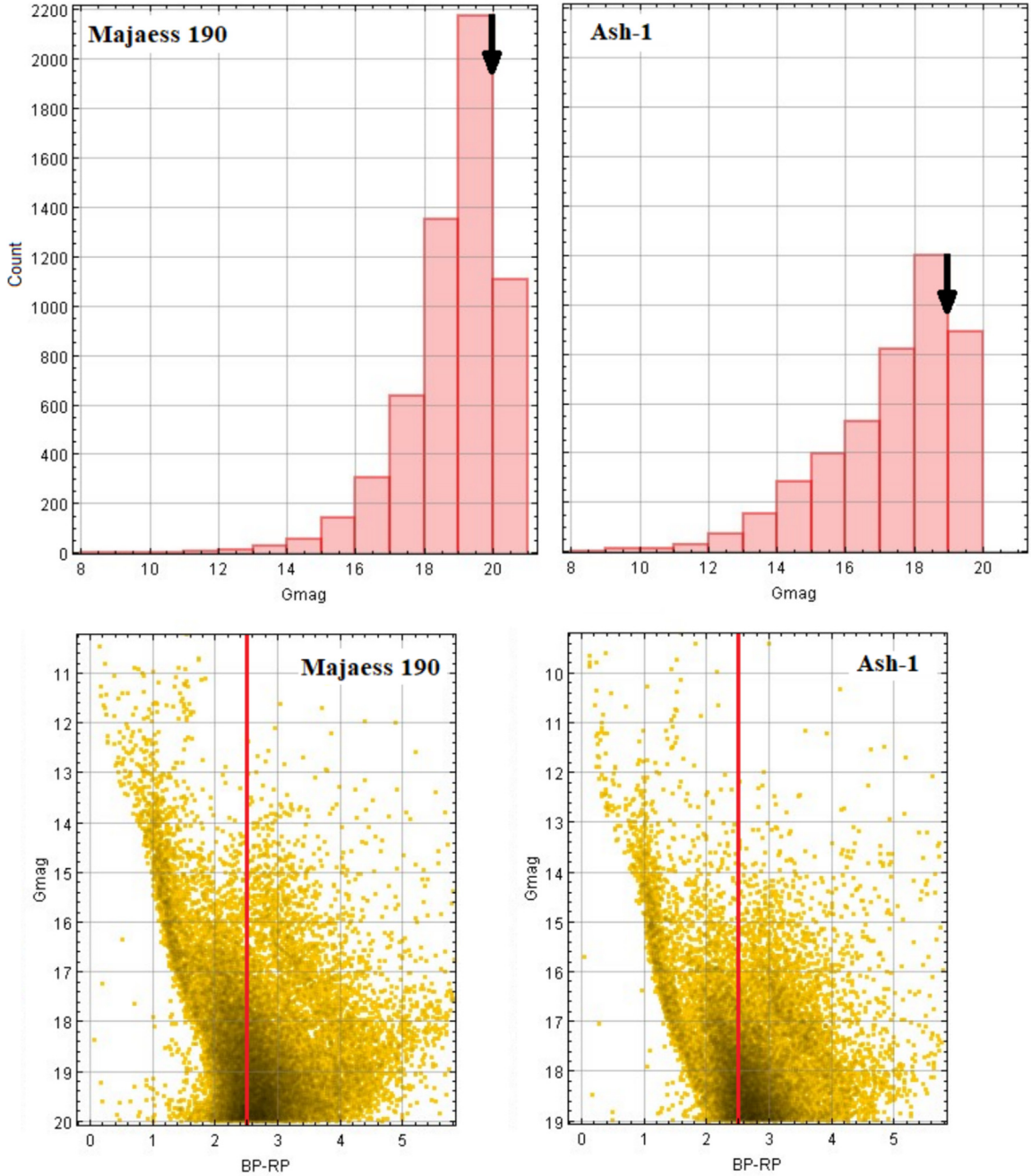

Fig. 2. The upper panels show G-band magnitude histograms for Majaess 190 and Ash-1, to exclude the very faint stars, where the arrows refer to the completeness limiting magnitudes, which are taken to be 20.0 mag and 19.0 mag for Majaess 190 and Ash-1, respectively. The lower panels represent the color cuts of (BP − RP) < 2.5 mag, to exclude the highly reddened stars; shown by red lines.

The radial stellar density profile of Ash-1 shows an approximately exponential decline in the surface density from the center outward, i.e., this overdensity acts as a star cluster. In addition, the stellar distribution in the color-magnitude diagram, consequently proved that it is a real star cluster. Finally, the pattern of its mass function and the amount of its relaxation time indicate that Ash-1 is a dynamically relaxed open cluster.

## 3. Data

The global website SIMBAD (https://simbad.u-strasbg.fr/simbad/) provides us with the central coordinates of star clusters. Majaess 190 lies in the constellation Sagittarius, at J2000.0 coordinates $\alpha = 18^h\,8^m\,20^s$, $\delta = -22°\,4'\,32.02''$, $\ell = 8.50381°$, $b = -0.98363°$. The coordinates of Ash-1, that we have newly discovered, are $\alpha = 18^h\,4^m\,59^s$, $\delta = -22°\,10'\,47''$, $\ell = 8.03460°$, $b = -0.35703°$.

With radii of 20 arcmin, the source data were downloaded from the main Gaia DR3 database via the VizieR website (https://vizier.cds.unistra.fr/viz-bin/VizieR-3?-source = I/355/gaiadr3). Fig. 1 displays the two clusters' optical images as obtained from ALADIN's colored DSS2 picture (https://aladin.u-strasbg.fr/AladinLite), where the fields of view (FOV) for both objects are 21.12 × 13.75 arcmin. The lower panel of Fig. 1 shows the field that includes the locations of the two clusters with the coordinate grid. We achieved 46,272 stars for Majaess 190 and 35,183 stars for Ash-1. The photometric completeness limitations have been examined for excluding the very dim stars. G-magnitude histograms were created for every cluster, as seen in the upper panels of Fig. 2. Reduced stellar counts for magnitudes lower than G = 20.0 mag for Majaess 190, and 19.0 mag for Ash-1. Also, to exclude the highly reddened stars, we performed color cuts of (BP − RP) < 2.5 mag for both clusters, as shown in the lower panels of Fig. 2. Evaluations did not include stars that were fainter and reddened than those limits. We plot the errors in parallax and PM components with the G-band for Majaess 190 and Ash-1, as shown in Fig. 3. We found that for the stars with completeness limiting magnitudes, the mean internal errors are ∼1.0 mas in Plx, and ∼1.0 mas yr$^{-1}$ in PM for both clusters.

The exact PM is sufficient to provide an initial selection of cluster members. We create the vector point diagram (VPD) for both clusters, where the distribution of the cluster stars is centered around a common point. We used the Virtual Observatory software TOPCAT, which handles astronomical tables very well, Tadross (2018). The VPD is displayed as shown in the left-hand panels of Fig. 4. A limited section in the darkest region of the VPD is chosen as the initial group (clump or subset) that has the most likely member stars, Bragaglia (2018), those who have the same mean PM of the cluster, (Sariya 2021); because of the significant of the dispersion in proper motion, see Cantat-Gaudin & Anders (2020). They have been tested on the color-magnitude diagram (CMD), where the stars that are close to the main sequence traces have been nominated to be cluster members as well, as shown in the middle panels of Fig. 4. The right-hand panels of Fig. 4 represent the proper motion components for each cluster's clump and the corresponding field background in the same respective area, respectively. Moreover, if the renormalized unit weight error (RUWE) is less than 1.4, it suggests that the single-star model fits the astrometric measurements well. Therefore, all stars that have RUWE greater than 1.4 were excluded.

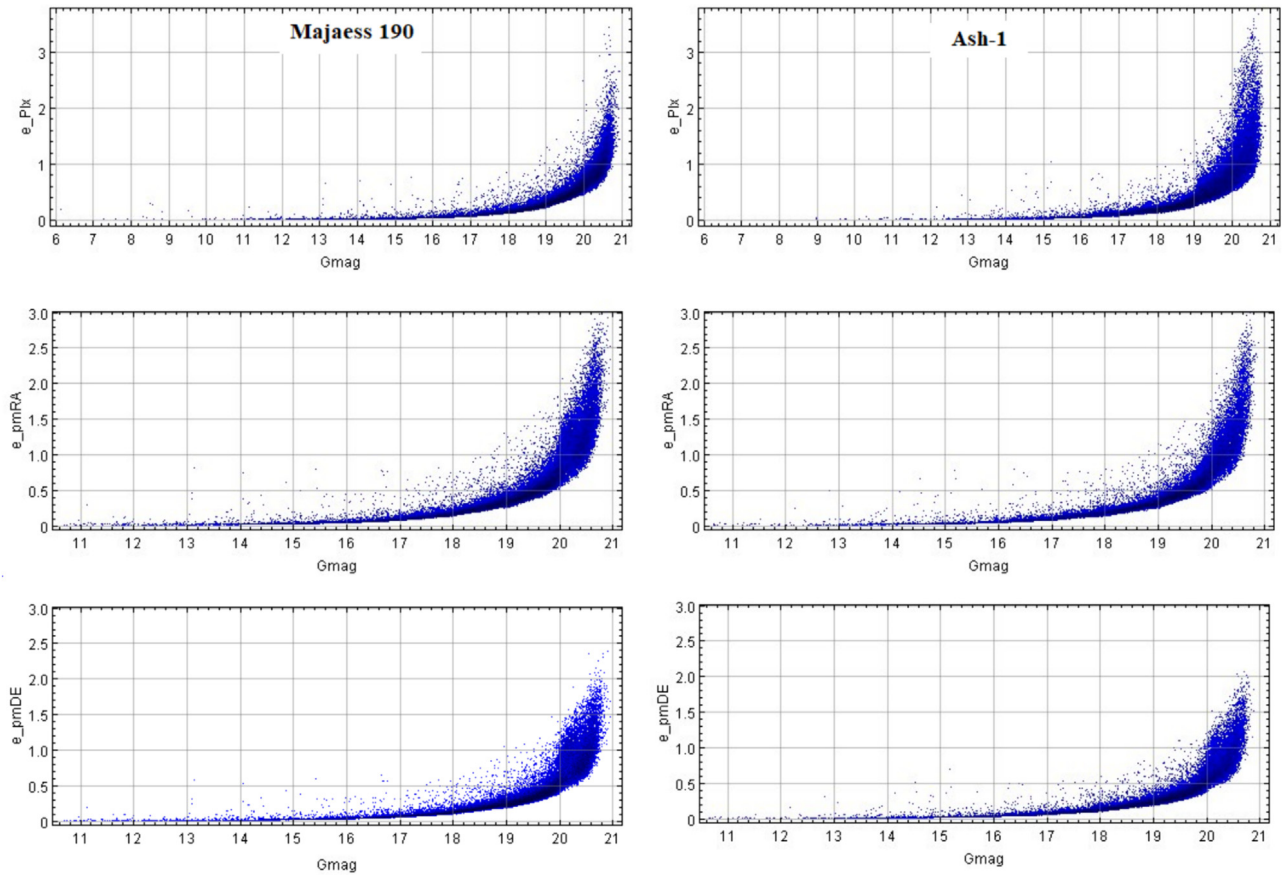

Fig. 3. The parallax errors (e_Plx mas), and the errors in the proper motion components (e_pmRA & e_pmDE mas/yr) are plotted against G-magnitude for Majaess 190 and Ash-1. The mean internal errors are ∼ 1.0 mas in Plx, and ∼ 1.0 mas yr$^{-1}$ in PM for both clusters.

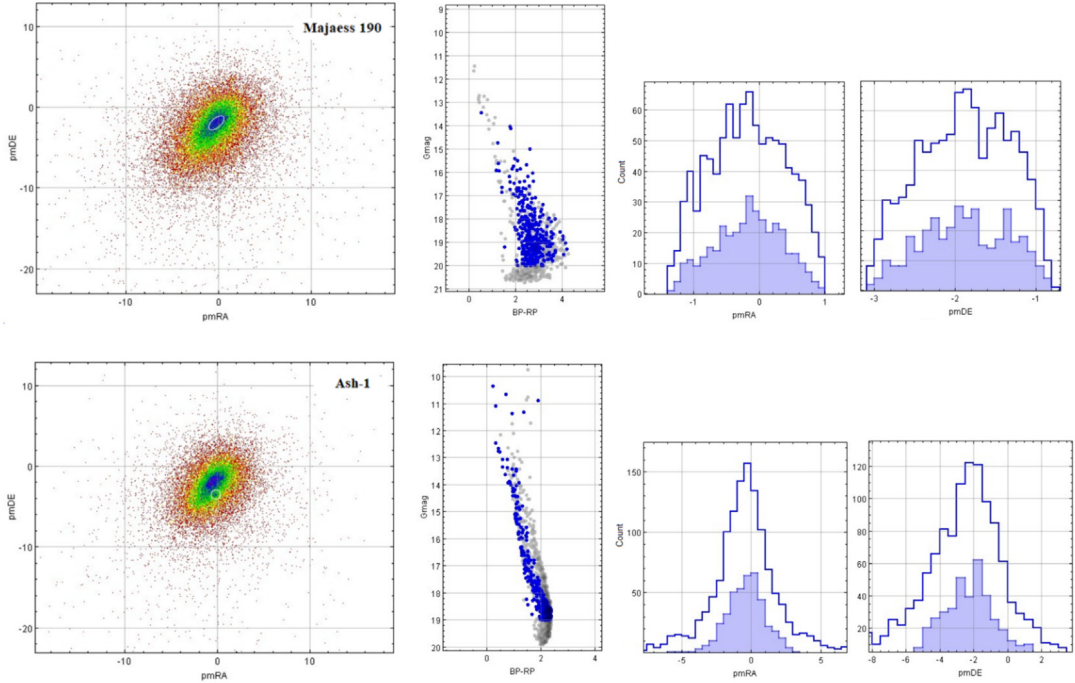

Fig. 4. The left-hand panels show the vector point diagrams of the stars in Majaess 190 and Ash-1. The densest areas are chosen as initial subsets (clumps) that contain the most likely members of the clusters. The impact and test of such clump's stars on the CMDs are displayed in the middle panels. The right-hand panels represent the proper motion components for the clump of each cluster (shaded area) and the corresponding field background in the same respective area.

The relation between the magnitudes and parallaxes of the two clusters is depicted in the left-hand panels of Fig. 5, where the clusters' stars are condensed, forming a peak adjacent to the field stars' peak. The right-hand panels display the histograms of parallaxes of the chosen clumps of Majaess 190 and Ash-1. Lindegren et al. (2021) reported a code for the zero-point offset of the Gaia EDR3 parallaxes (https://gitlab.com/icc-ub/public/gaiadr3_zeropoint). The zero-point correction was approximated as a function of G-magnitude, ecliptic latitude, and color (using $\nu_{eff}$ for the five-parameter solutions and the pseudo-color for the six-parameter solutions). Applying this code to the likely members of each studied cluster's clump. The corrected parallaxes are found to be 0.469 ± 0.09 and 0.736 ± 0.07 mas for Majaess 190 and Ash-1, respectively. These lead to derived distances of ∼ 2130 and 1360 (±80) pc, respectively.

4. Clusters' dimensions

To estimate the real cluster center, the cluster's area has been divided into equal-sized bins in right ascension (RA) and declination (DE), and the stars in each bin have been counted. The cluster center is located in the densest part of the cluster area. The Gaussian fitting profile has been applied in both directions, as shown in Fig. 6. The estimated central coordinates of Majaess 190 and Ash-1 are found to be shifted by 15.3$^s$ & 3.44$^s$ in RA and 26.08″ & 8.2″ in DE, respectively. However, the uncertainties in position ($\ell$, b) for both clusters are found to be trivial values and can be neglected. On the other hand, to estimate the cluster radius, we split the cluster's region into equal concentric rings (shells), i.e., every 0.5 arcmin from the estimated cluster center. The radial density profile (RDP) can be obtained by calculating the stellar density of each shell (dividing the total star number at that shell by its surface area, Tadross (2023); Tadross & Elhosseiny (2022). The density profile frequently shows an approximately exponential drop in the surface density of the cluster from the center out. The radius at which the surface density of the cluster roughly equals the mean density of the surrounding field is known as the limited radius ($R_{lim}$), Joshi et al. (2012). It is assumed that each shell's density errors follow the Poisson noise ($1/\sqrt{N}$), where N is the total number of stars in that shell, see Fig. 7. According to King (1966), King's model is applied to the studied clusters as follows:

$$f(R) = f_{bg} + \frac{f_0}{1 + \left(\frac{R}{R_c}\right)^2}$$

where $f_{bg}$, $f_o$ are the background, and central density respectively. $R_c$ is the core radius of the cluster, which is defined as the distance at which the star density reduces to half of the central density ($f_o$). After a certain point, where the density of the cluster dissolved into the surrounding field, the RDP reduced and then stabilized. $R_{lim}$ can be computed as follows using Bukowiecki's (2011) formula:

$$R_{lim} = R_c \sqrt{\frac{f_0}{3\sigma_{bg}} - 1}$$

where $\sigma_{bg}$ is the uncertainty of the $f_{bg}$ value. Keep in mind that the estimated radius indicates the radial minimum limit of the cluster, Maurya & Joshi (2020). Hence, the estimated radii $R_{lim}$ and core radii $R_c$ are found to be 6.0, and 11.0 (±0.25) arcmin; and 0.76 and 1.81 (±0.11) arcmin for Majaess 190 and Ash-1, respectively. Conversely, the tidal radius ($R_t$) is defined as the separation from the cluster center at which the Galaxy's gravitational pull equals that of the cluster core. Using Jeffries (2001)'s equation, we can get the tidal radius as follows:

$$R_t = 1.46 \times M_c^{\frac{1}{3}}$$

where $R_t$ is the cluster's tidal radius, measuring in parsecs, and $M_c$ is the total mass of the cluster, measuring in solar units (see Sec. 5). $R_t$ has been estimated in arcmin and found to be 15 and 25 arcmin for Majaess 190 and Ash-1, respectively (see Fig. 7). As mentioned by Peterson & King (1975), $C = \log (R_{lim}/R_c)$ is the concentration parameter, which reveals how the cluster contrasts with the nearby field stars. It turns out to be 0.90 and 0.78 for Majaess 190 and Ash-1, respectively.

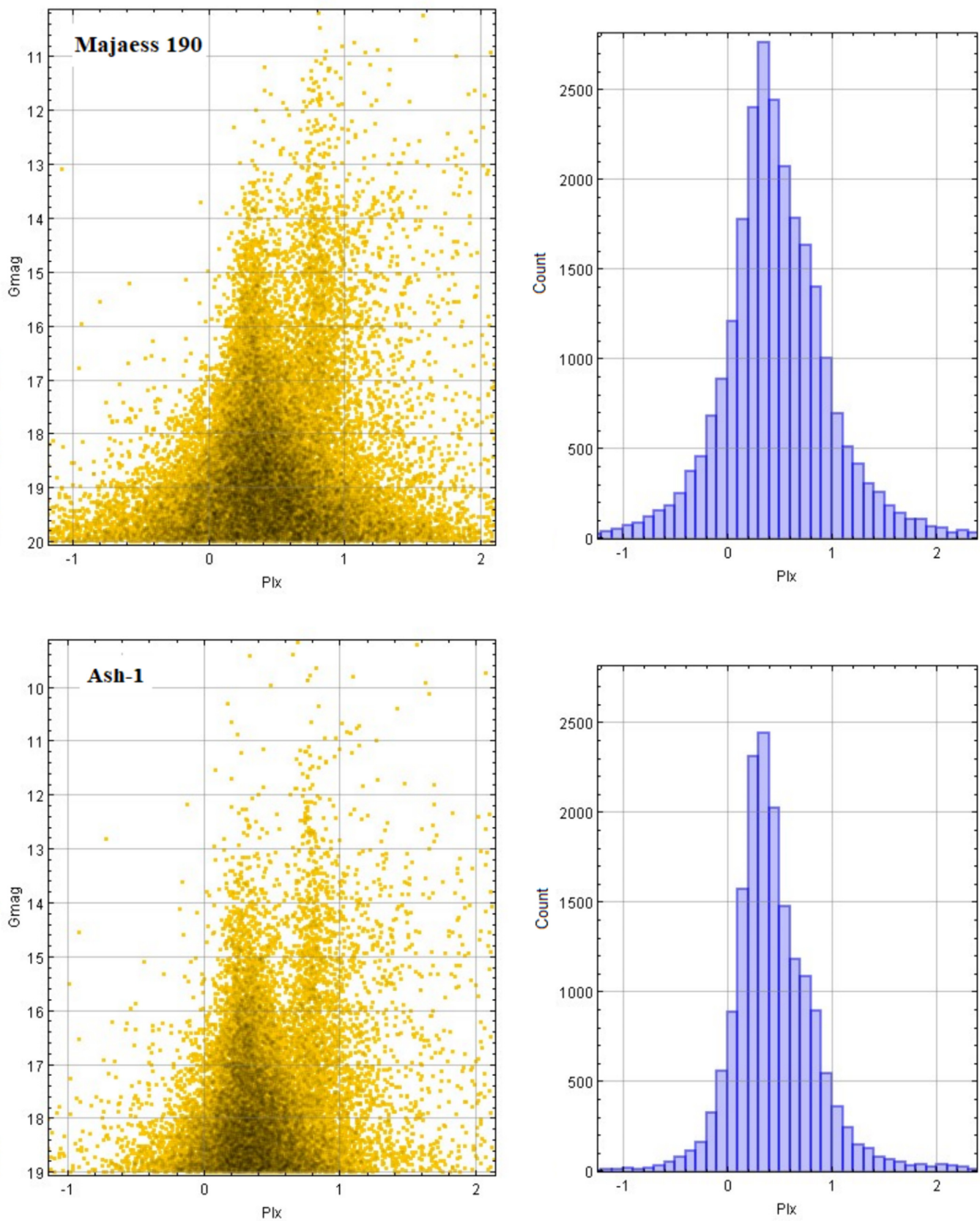

Fig. 5. The left-hand panels show the correlation between the magnitude and parallax of each cluster. The cluster's stars seem to be concentrated, forming a peak adjacent to the peak of the field stars. The right-hand panels show the parallaxes histograms of the selected clumps of Majaess 190 and Ash-1.

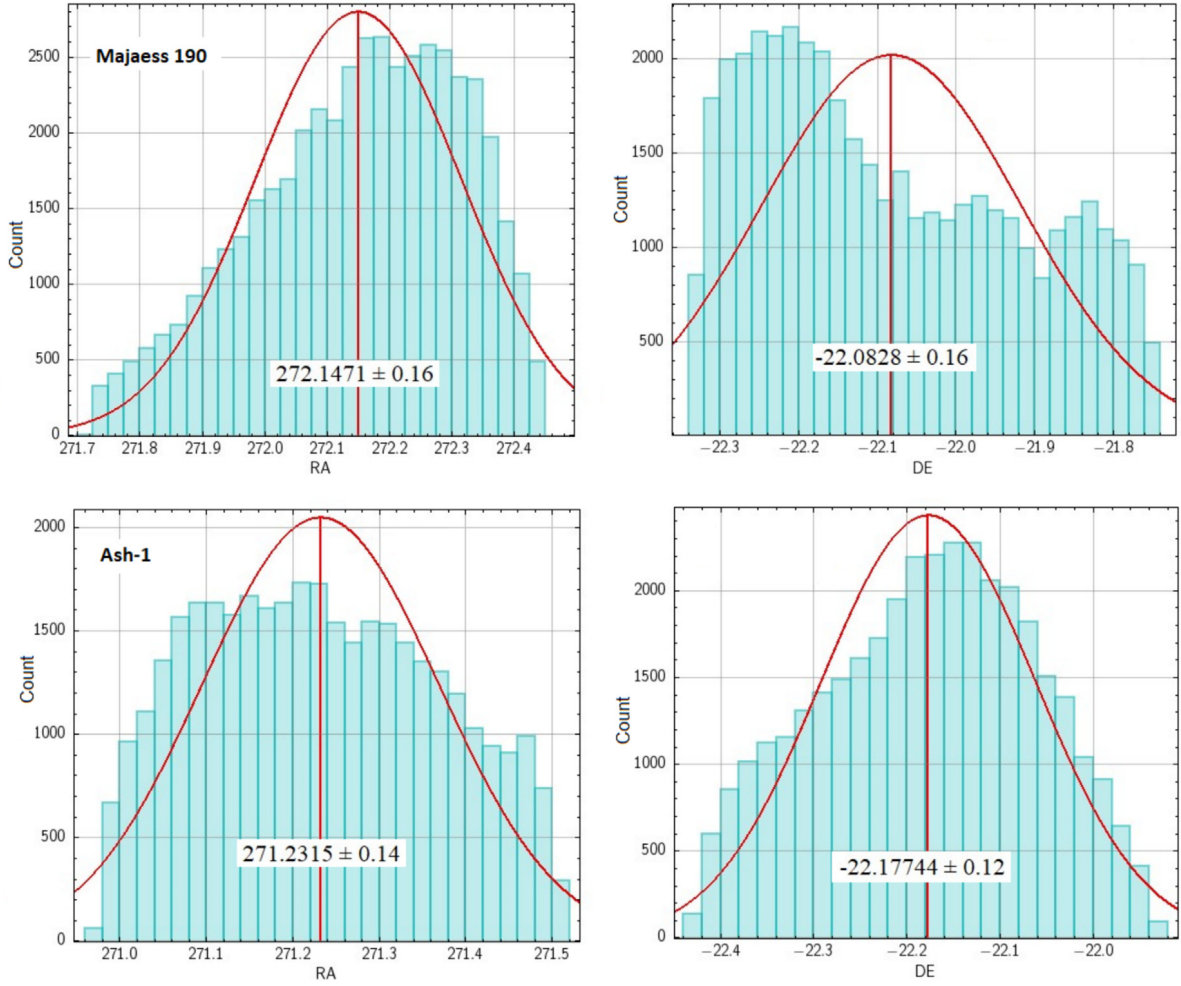

Fig. 6. Centers estimations of Majaess 190 and Ash-1. The Gaussian fitting profiles are applied to RA and DE. The estimated current centers are noticed to be shifted by 15.3$^s$ & 3.44$^s$ in RA, and 26.08″ & 8.2″ in DE, respectively.

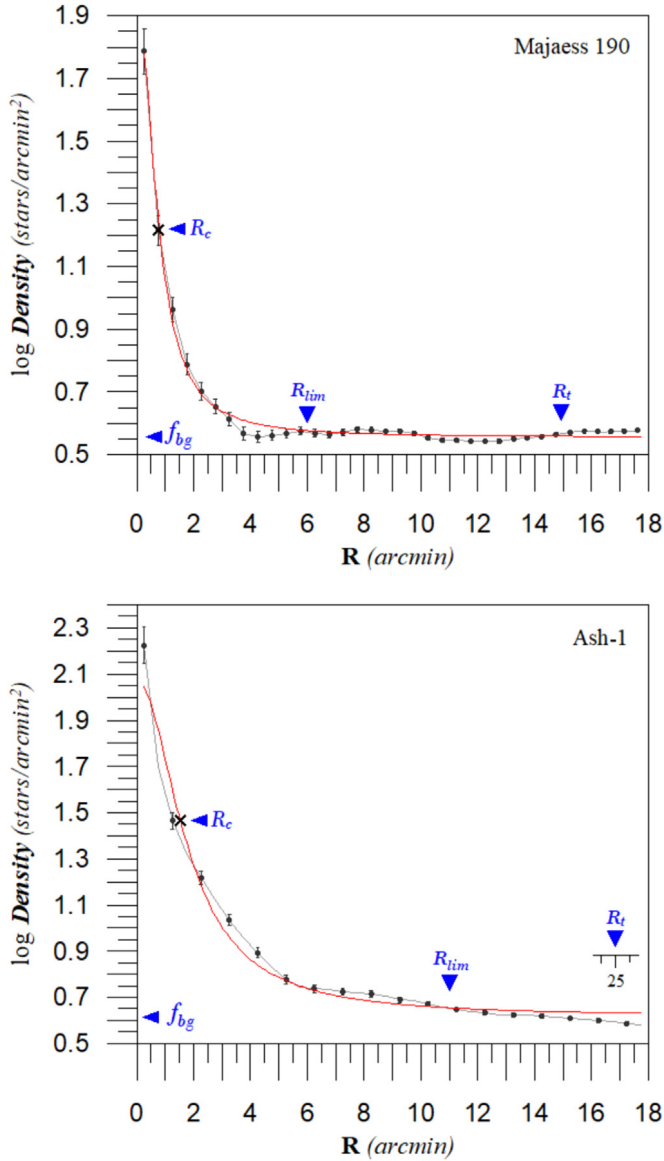

Fig. 7. Radial density profiles of Majaess 190 and Ash-1. The red line represents the fitting of King's (1966) profile, while the error bars refer to the Poisson distribution $1/\sqrt{N}$. The blue triangles refer to the background field star density $f_{bg}$, core radii $R_c$, the limited radii $R_{lim}$, and the tidal radii $R_t$ of the two studied clusters, respectively.

## 5. Membership candidates

The pyUPMASK algorithm (Pera et al. 2021) was used to assign the membership probabilities for the cluster stars. This algorithm (https://github.com/msolpera/pyUPMASK) is a widely used open-source software program that was created using the Python programming language to identify member stars according to their photometric and astrometric properties, Cantat-Gaudin & Anders (2020). UPMASK (Unsupervised Photometric Membership Assignment in Stellar Clusters) accesses five parameters (RA, DE, pmRA, pmDE, and Plx) of Gaia DR3 data. The most likely members were considered to be the stars with a probability P > 0.50. Furthermore, the star is classified as a member, if its 3σ of parallax matches the mean value of the cluster, situated near the main sequence curve, i.e., lies in a slice around the main sequence not exceeding 0.15 mag on both sides, and at the same time, lies in the estimated limited radius of the cluster (effective radius). We obtained 355 and 410 members candidates for Majaess 190 and Ash-1, respectively, see Fig. 8.

## 6. Color-magnitude diagrams

The color-magnitude diagram (CMD) is a crucial tool for determining the essential features of the cluster stars. Using the Padova stellar isochrones database of PARSEC (https://stev.oapd.inaf.it/cgi-bin/cmd), the stellar evolutionary isochrones and tracks of Bressan et al. (2012) are used with the Gaia filter pass-bands of Evans et al. (2018) for solar metallicity Z = 0.0152. It is possible to estimate the color excess, metallicity, age, and distance modulus simultaneously by finding the best fit for the cluster's CMDs. Based on our visual fit, as illustrated in Fig. 9. The cluster's age error is determined when the cluster's main sequence lies between two ages of standard isochrones. Correspondingly, the distance modulus error is estimated according to the best fitting of the main sequence in the two previous cases. The photometric fundamental characteristics of Majaess 190 and Ash-1 are proven to be 4 ± 0.05 Gyr and 630 ± 20 Myr, respectively. The distance moduli and color excess are identified to be 11.70, 10.68 (±0.10) mag, and 1.20, 0.40 (±0.12) mag, respectively. These results agree rather well with our mean parallax values. Fig. 9′s right side displays the cluster stars' membership probability with the color scales of each cluster. According to the CMD-3.6 input form (https://stev.oapd.inaf.it/cgi-bin/cmd), the passbands of the Gaia filters G, $G_{BP}$, and $G_{RP}$ are the following:

| Filter | G | $G_{BP}$ | $G_{RP}$ |
|---|---|---|---|
| $A_\lambda / A_v$ | 0.83627 | 1.08337 | 0.63439 |

we get $A_{GBP} = 1.083\ Av$, $A_{GRP} = 0.634\ Av$, and $A_G = 0.836\ Av$, then $Av = 2.227\ E(G_{BP} - G_{RP})$ and $A_G = 1.862\ E(G_{BP} - G_{RP})$. By converting the color excess to $E(B - V)$ and correcting the magnitudes for interstellar reddening, these ratios have been employed with $Av = Rv\ E(B - V)$, where $Rv = 3.1$, Zhong et al. (2019). For Majaess 190 and Ash-1, the related values of E(B-V) are computed to be 0.91 and 0.30 (±0.10) mag, respectively.

To ensure more precision, we applied the ASteCA code (https://asteca.readthedocs.io/en/latest/) to the studied clusters, Perren et al. (2015, 2020). The generated values of the ASteCA code agree fairly well with ours. Furthermore, the 3D extinction map (https://argonaut.skymaps.info) and Stilism (https://stilism.obspm.fr) at the clusters' positions displayed that their reddening and distances accord well with our estimates as well. According to Tadross (2011), the distances from the galactic center ($R_g$), where $R_o = 8.2$ kpc, Bland and Gerhard (2016), and the Cartesian galactocentric positions $(X_\odot - Y_\odot - Z_\odot)$ are estimated for the two studied clusters. Recognizing that the X-axis is perpendicular to the Y-axis, which is positive in the first and second Galactic quadrants, and that the Y-axis connects the Sun to the Galactic center, which is positive toward the Galactic anti-center, Lynga (1982), see Table 2.

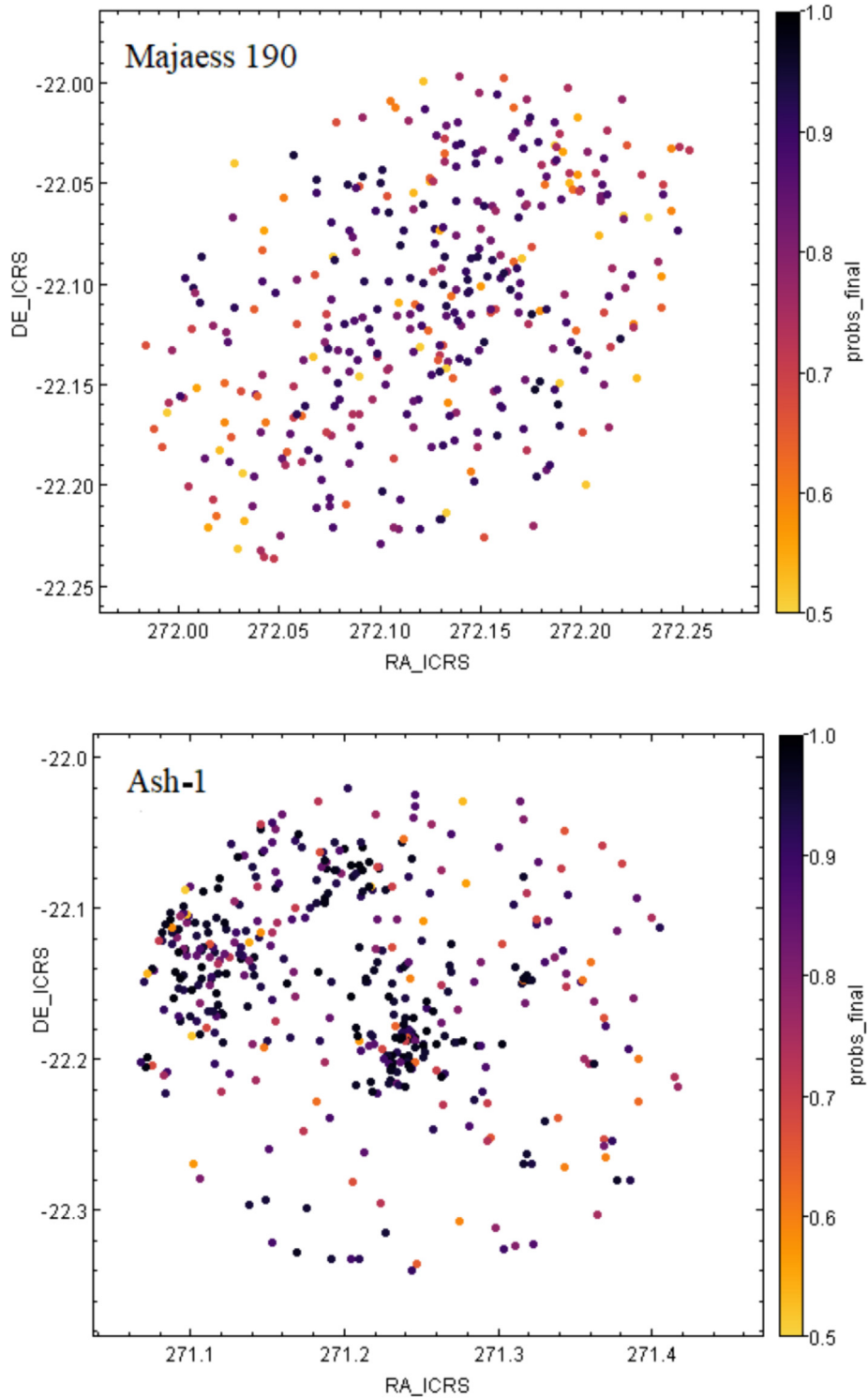

Fig. 8. The distribution map of the member stars of Majaess 190 and Ash-1 using the equatorial position coordinates (RA, DE) of Gaia data. The membership probabilities of the stars are shown according to the color scales on the right side of the figure.

## 7. Mass, luminosity functions, and dynamic state

With the likely member stars, we can achieve the mass and luminosity functions (MF & LF) for Majaess 190 and Ash-1. In a nutshell, MF is the count of stars per logarithmic mass interval while LF is the count of stars per absolute magnitude interval. Keep in mind that, cleaning

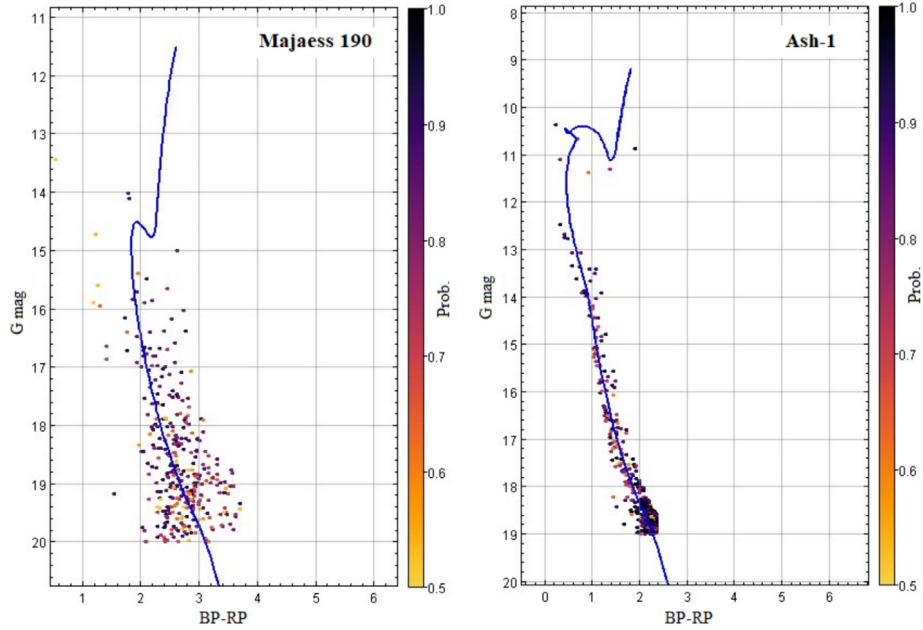

Fig. 9. CMDs of the likely member stars of Majaess 190 and Ash-1 fitted with theoretical Padova isochrones of ages 4 ± 0.05 Gyr and 630 ± 20 Myr, respectively. The membership probabilities of the stars are shown according to the color scales on the right side of the figure, where P > 0.50. The distance moduli and reddening are found to be 11.70 & 10.68 (±0.10) mag, and 1.20 & 0.40 (±0.12) mag, respectively.

Table 2
The astrophysical properties of the studied clusters.

| Parameter | Majaess 190 | Ash-1 |
| --- | --- | --- |
| $\alpha$ (2000) h:m:s | 18: 08: 35.80 | 18: 04: 55.56 |
| $\delta$ (2000) °: ′:″ | –22: 04: 58.20 | –22: 10: 38.80 |
| $\ell$ (2000) deg | 8.50380 | 8.03460 |
| $b$ (2000) deg | −0.98363 | −0.35704 |
| Age Myr | 4000 ± 50 | 630 ± 20 |
| $E(B_P-R_P)$ mag | 1.20 | 0.40 |
| $E(B-V)$ mag | 0.91 | 0.30 |
| $A_G$ mag | 2.23 | 0.74 |
| $A_V$ mag | 2.83 | 0.94 |
| $(m-M)_o$ mag | 11.70 ± 0.10 | 10.68 ± 0.10 |
| Dist. pc | 2140 ± 100 | 1350 ± 100 |
| Plx. mas | 0.469 ± 0.09 | 0.736 ± 0.07 |
| pmRA mas/yr | −0.17 ± 0.14 | −0.37 ± 0.11 |
| pmDE mas/yr | −1.87 ± 0.09 | −2.18 ± 0.08 |
| Mem. stars | 355 | 410 |
| $f_{bg}$ | 0.55 | 0.62 |
| $f_o$ | 1.36 | 1.45 |
| $R_{lim}$ arcmin | 6.0 (3.7 pc) | 11.0 (8.8 pc) |
| $R_c$ arcmin | ∼ 0.76 | ∼ 1.81 |
| $R_t$ arcmin | ∼ 15.0 | ∼ 25.0 |
| C | 0.90 | 0.78 |
| $R_g$ kpc | 6.09 | 6.87 |
| $X_\odot$ pc | −2114 | −1336 |
| $Y_\odot$ pc | 316 | 188 |
| $Z_\odot$ pc | −36.7 | −8.4 |
| T. Lum. mag | −0.70 | −3.2 |
| T. mass $M_\odot$ | 375 | 455 |
| $T_R$ Myr | 19 | 25 |
| $\tau$ | 207 | 26 |

field star contamination from the cluster region is the difficult part of analyzing the LF and MF.

The unique distance modulus of each cluster is used to translate the apparent magnitudes of the cluster members into absolute magnitudes. At the age of each cluster, we used the main sequence stars of the isochrones that fit the cluster CMD to generate a polynomial equation that aided us in calculating each member's mass and luminosity. The theoretical tables of evolutionary tracks of Marigo et al. (2017) have been used. By adding the results of multiplying the count members in each bin by the values of that bin, the total mass and total luminosity of the cluster will be obtained (Tadross, 2012). Fig. 10 shows one histogram for each cluster, where the MF and LF are merged. It represents the relation between the logarithmic count members and the logarithmic mass intervals, the upper axis shows the absolute magnitude intervals. It also refers to mass segregation, in which the brightest, most massive members are concentrated in the cluster's core, while the fainter and less massive members are dispersed outside. In our cases, the total luminosity values are found to be –0.47 and –2.11 mag for Majaess 190 and Ash-1, respectively. The linear fit depicts the initial mass function (IMF) slope of the equation as follows:

$$log \frac{dN}{dM} = -\alpha \, log\,(M) + const.$$

where $dN/dM$ is the count of members in the mass interval $[M:(M+dM)]$ and $\alpha$ is the slope's value of the relation, where its standard mean value is ≈ 2.35, according to Salpeter (1955). The overall masses are found to be 250 $M_\odot$ and 303 $M_\odot$ for Majaess 190 and Ash-1, respectively.

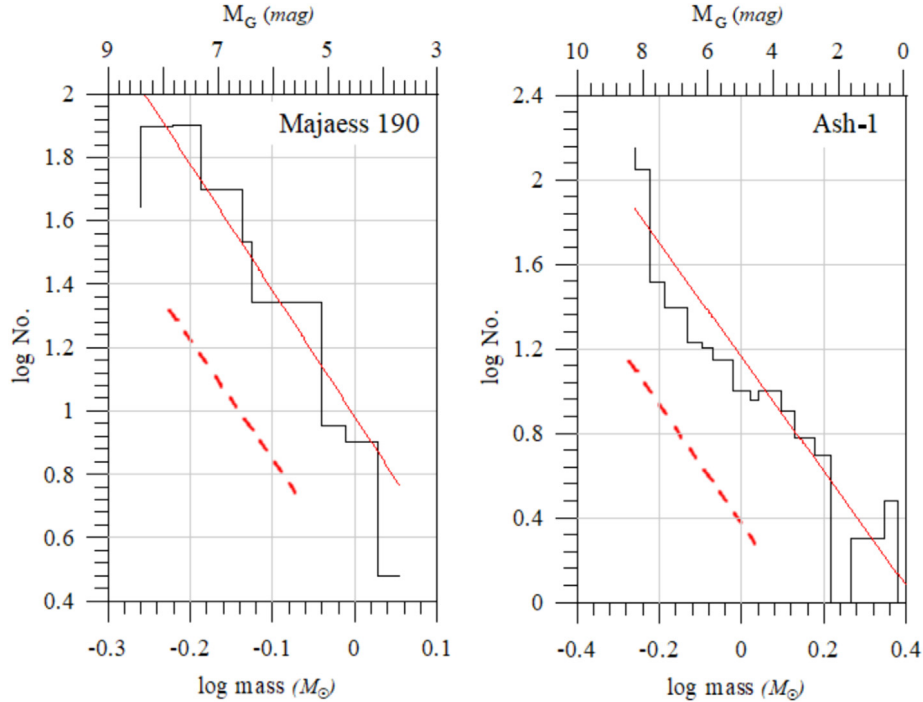

Fig. 10. The mass and luminosity functions' histograms of the clusters under study. MF represents the relation between the count members and the logarithmic mass intervals, while the upper axis represents the absolute magnitude intervals. The mass function slopes of Majaess 190 and Ash-1 are shown by the solid red lines, which coincide with the mean value reported by Salpeter (1955), the dashed red lines.

The impact of unresolved binaries was examined by Piskunov & Malkov (1991, and references therein), where the mass derived from the mass-luminosity relation is less than the total mass of the individual components, which is called the mass defect. They concluded that about 50 % of all main sequence stars are in multiple systems. Recently, Giacomo et al. (2023, and references therein) studied the unresolved binary systems in 78 Galactic open clusters, using Gaia DR3, where each consists of two main sequence stars. They concluded that the fraction of binary stars ranges from ∼15 % to more than ∼60 % without a discernible relationship between the mass and the age of the host cluster. Therefore, according to the influence of unresolved binaries, the LF and MF could be increased by 50 %. So, the total luminosity values would be –0.70 and –3.2 mag for Majaess 190 and Ash-1, respectively. Also, the estimated total mass would be 375 $M_\odot$ and 455 $M_\odot$ for Majaess 190 and Ash-1, respectively.

Conversely, the dynamic state of each cluster can be estimated from its relaxation time ($T_R$). It is defined as the amount of time that the cluster needs from the start to self-assemble and reach a stable state in opposition to the pressures of contraction and destruction. It is also described as the time interval within which the cluster completely loses all evidence of its early conditions, mostly based on the cluster's total mass and the count of potential member stars. However, $T_R$ is the normal time scale that the cluster needs to reach equilibrium energy. It can be computed using the formula of Spitzer and Hart (1971) as follows:

$$T_R = \frac{8.9 \times 10^5 \sqrt{N} \times R_h^{1.5}}{\sqrt{\langle \bar{m} \rangle} \times \log(0.4N)}$$

where $N$ is the count of the cluster's members, $R_h$ (in parsecs) is the radius that includes half of the cluster's total mass, and $\bar{m}$ (in solar unit) is the mean mass of a cluster's member. We first summed up the masses of each member and computed the cluster's total mass, and then we got the mean mass of a member. Hence, we can determine the $R_h$ value at the radius where half of the cluster's total mass is included. From there, we can determine the $T_R$ value, which for Majaess 190 and Ash-1 is found to be 19 and 25 Myr, respectively.

The dynamical evolution parameter can be defined as $\tau = Age/T_R$, where the age of the cluster should be greater than the relaxation time, i.e. $\tau \gg 1.0$, thus, the cluster can be said to be dynamically relaxed, and vice versa. When the ages of the clusters are compared to their relaxation times, it is found that the relaxation times are significantly smaller than the cluster ages; $\tau = 207$ and 26 for Majaess 190 and Ash-1 respectively. This implies that the clusters under study are certainly dynamically relaxed.

## 8. Conclusions

Accurate membership determination is an essential process for astrophysical studies of open star clusters, and it depends mainly on the quality of the data. Thanks to the Gaia DR3 database, which improved our ability to specify the memberships of star clusters and calculate their main physical characteristics. We investigated here a comprehensive study of two open star clusters, the first is a poorly studied cluster in the constellation Sagittarius (Majaess 190), which has never been completely investigated. The second one was accidentally discovered in the field of the first (named Ash-1), which is located very close to the newly discovered system CWNU-2598 (by about 0.05 deg only). We believe that the two clusters are different with some overlapping because they form a binary system, which requires another detailed study. The main astrophysical characteristics of Majaess 190 and Ash-1 have been estimated here for the first time. The membership probabilities of stars were assigned using the pyUPMASK algorithm. The parallaxes offset has been corrected using the zero-point correction code of Lindegren et al. (2021). Also, we employed the ASteCA code to ensure our values of the estimated parameters. Ages, extinctions, and heliocentric distances are estimated. The mass function, luminosity function, and total masses were determined as well. In addition, on studying the dynamic state, we found that Majaess 190 and Ash-1 are dynamically relaxed clusters. The main physical parameters identified by this investigation are listed in Table 2.

**Declaration of competing interest**

The authors declare that they have no known competing financial interests or personal relationships that could have appeared to influence the work reported in this paper.

**Acknowledgement**

This work made use of data from the European Space Agency (ESA) mission Gaia (https://www.cosmos.esa.int/Gaia), processed by the Gaia Data Processing and Analysis Consortium (DPAC, https://www.cosmos.esa.int/web/Gaia/dpac/consortium). Funding for the DPAC was provided by national institutions, in particular, the institutions participating in the Gaia Multilateral Agreement. This work made use of the SIMBAD database and the VizieR catalog access tool, operating at the CDS, Strasbourg, France (DOI: 10.26093/cds/vizier), and of NASA Astrophysics Data System Bibliographic Services.